\documentstyle[12pt,a4wide,amsmath,amssymb,epsf]{article}
\newcommand{\nin}{\noindent}
\newcommand{\be}{\begin{equation}}
\newcommand{\ee}{\end{equation}}
\newcommand{\bea}{\begin{eqnarray}}
\newcommand{\eea}{\end{eqnarray}}

\newcommand{\nn}{\nonumber\\}

\begin{document}
                                                                                                                    
\begin{center}
                                                                                                                    
{\Large{\bf Concepts of Renormalization in Physics}}
                                                                                                                    
\vspace{1cm}
                                                                                                                    
{\bf Jean Alexandre}\\
Physics Department, King's College \\
WC2R 2LS, London, UK\\
jean.alexandre@kcl.ac.uk
                                                                                                                    
\vspace{1cm}
                                                                                                                    
{\bf Abstract}
                                                                   
\end{center}
                                                                                                                    
\vspace{.2cm}
                                                                                                                    
A non technical introduction to the concept of renormalization is given,
with an emphasis on the energy scale dependence in the description of a physical system.
We first describe the idea of scale dependence in the study of a ferromagnetic phase transition,
and then show how similar ideas appear in Particle Physics.
This short review is written for non-particle physicists and/or students aiming at studying 
Particle Physics. 

\vspace{1cm}

\section{Introduction}

Nature provides us with a huge amount of phenomena, which can be explained by different 
theories, depending on the scale of the physical processes. At the atomic level: Quantum Mechanics
is the most relevant theory; in every-day life: Newtonian Mechanics explains the trajectories
of rigid bodies; at the cosmological scale, General Relativity is necessary to describe the
evolution of the Universe. 

Obviously, to each scale corresponds another set of parameters that help us describe Physics. 
This is actually valid within a given theory and one does not need to change the scale of observation
so dramatically to see different descriptions of the system:
consider the example of fluid mechanics.
The Reynolds number of a flow is a characteristic dimensionless quantity which helps define
different regimes. It is given by $R_e=UL/\nu$, where $\nu$ is the viscosity of the fluid, $U$ and $L$ are typical
speed and length of the flow respectively. Suppose that $U$ and $\nu$ are fixed.
For fluids flowing over short distances ($R_e<<1$), viscosity effects dominate and inertia is negligible.
Surface tension can also play a role. For fluids flowing over large
distances ($R_e>>1$), viscosity can be neglected and non linearities dominate the system, leading to
turbulence and instabilities. Therefore we see that, depending on the typical scale over which the
fluid is observed, different parameters have to be considered to describe its flow.
 
In a very general manner, "renormalization" deals with the evolution of the description of a system 
with the scale of observation.
Renormalization was introduced as a tool to predict physical properties in phase transitions, 
as will be described in this article, and Kenneth Wilson was given for this the Nobel Prize in 1982
(seminal papers of his are \cite{wilson}). 
Renormalization also happens to be necessary to avoid mathematical inconsistencies when computing
physical quantities in Particle Physics. 
Historically though, renormalization appeared in Particle Physics 
independently of its use for the description of phase transitions, but it was then understood 
that both procedures actually have the same physical content, and a unifying description of 
renormalization was set up. This introductory review aims at describing these ideas.  

In section 2, we start by an intuitive description of the technique which leads to 
the idea of scale dependence, in a ferromagnetic system. We explain here how to obtain 
the temperature dependence of physical quantities, near a phase transition.
The appearance of scale dependence in Particle Physics
is explained in section 3. We describe here how the necessity to introduce a regularization 
for the computation of quantum corrections to physical processes leads to a scale-dependent theory.
Section 4 comes back to the ideas developed for a ferromagnetic system, applied to Particle Physics,
and the connection between quantum fluctuations and thermal fluctuations is emphasized.

For non-physicist readers, a few concepts used in the text are defined in the appendix.

\section{Renormalization and phase transitions}

The use of scale dependence in the description of a system proved to be very fruitful in
predicting the behaviour of physical quantities, as magnetic susceptibility or
heat capacity, in the vicinity of a $2^{nd}$ order phase transitions (see def.1 in the appendix).
These ideas are explained in many places and good introductory books
are for example \cite{amit,lebellac}.

Consider a ferromagnetic sample at temperature $T$. This system is
made out of spins located at the sites of a lattice, and to each spin corresponds a magnetic moment
such that spins are related by a magnetic interaction.
When $T$ is larger than some critical temperature $T_C$, the magnetization of the sample is zero: 
thermal fluctuations are too important and destroy the magnetic order. When $T$ is less than $T_C$,
the spin interactions dominate the thermal fluctuations and spins are ordered along a given direction
(which depends on the past of the sample).

If we look at the system for $T$ well above $T_C$, each spin interacts mainly with its nearest neighbours:
the correlation length (the typical distance over which spins interact) is of the order of few lattice
spacings. But as $T$ approaches $T_C$, the magnetic interactions are
not dominated by thermal fluctuations anymore and play a more important role:
the correlation length grows. In this situation, a
fair description of the system must involve an increasing number of degrees of freedom and mean field 
approximations or naive perturbation expansions become useless. 

What saves us here is the following assumption (the "scaling hypothesis"):
the macroscopic behaviour of the system
depends on physical properties that occur at the scale of the correlation length $\xi$, and smaller scales
should not be responsible for it. The idea of what is called "renormalization group transformation"
(RGT) is then to get rid of
these smaller details by defining a new theory with lattice spacing $\xi$ instead of the
distance between the original lattice sites, so as to be left with a description in terms of 
relevant degrees of freedom only. An RGT thus has the effect to decrease the resolution in the
observation of the system.

\begin{figure}[ht]
\epsfxsize=12cm
\centerline{\epsffile{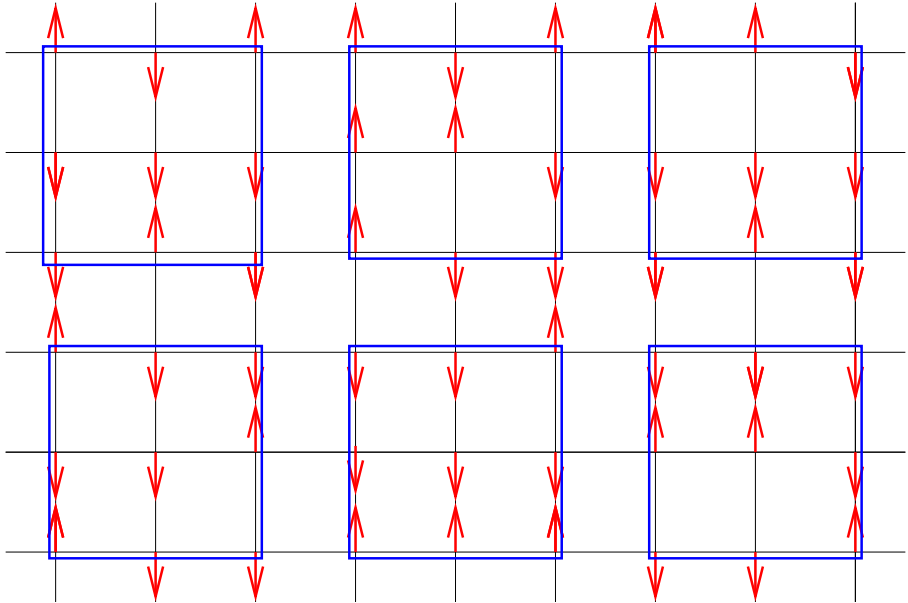}}
\caption{\it Example of a blocking procedure in a planar ferromagnetic system. The block spin variable 
$\sigma$ can for example be obtained by adding the values $\pm 1$ of the nine original spins 
inside the block: $\sigma=1$ if the sum is positive and $\sigma=-1$ if the sum is negative.}
\label{blocking}
\end{figure}

The procedure to achieve this is the following: starting from a system with lattice spacing $a$, 
one defines blocks of spins of size $sa$, where $s>1$ is the dimensionless scale factor (see fig.\ref{blocking}). 
To each 
block of spin corresponds a block variable, i.e. a new spin variable depending on the configuration
of the original spins inside the block. There are several ways to define this block variable and this
step corresponds to the so called blocking procedure or "coarse graining": 
the block variables describe the system as if the latter was zoomed out.
The new theory, defined on the lattice with spacing $sa$, is obtained by noting that 
the partition function (see def.3) of the system should be invariant under this coarse graining.
To see this more precisely, let $H$ be the Hamiltonian (see def.2) of the original system, 
defined on the original lattice with spacing $a$, and  
$H_s$ the Hamiltonian of the blocked system with lattice spacing $sa$. 
The invariance of the partition function $Z$ gives
\bea\label{Z1}
Z=\sum_{spins}\exp\left(-\frac{H}{k_BT}\right)=\sum_{blocks}\exp\left(-\frac{H_s}{k_BT}\right),
\eea
where $\sum_{spins}$ is the sum over the original spins and $\sum_{blocks}$ is the sum
over the block spins on the new lattice.
Eq.(\ref{Z1}) helps us define the block Hamiltonian $H_s$ as;
\bea\label{int1}
\exp\left(-\frac{H_s}{k_BT}\right)=\tilde\sum\exp\left(-\frac{H}{k_BT}\right),
\eea
where $\tilde\sum$ is the constrained sum over the original spins that leave a given configuration 
of the block spins unchanged.

Therefore one has a way to define the Hamiltonian $H_s$ of the coarse grained system, with
lattice spacing $sa$.
For each value of the scale factor $s$, $H_s$ is defined by a set of parameters $\{\mu_s\}$ and
the evolution of $\{\mu_s\}$ with $s$ constitutes the so called renormalization flows. 
These flows depend on the theory that is defined on the original lattice, as well as on the blocking procedure.
But the essential physical content lies in the behaviour of these flows at large distances: 
the different theories described by $H_s$, with running $s$, have the same large-scale behaviours, 
defined by the relevant parameters. Increasing $s$ 
removes the irrelevant microscopic details which do not influence the macroscopic physics.  
 
\begin{figure}[ht]
\epsfxsize=11cm
\centerline{\epsffile{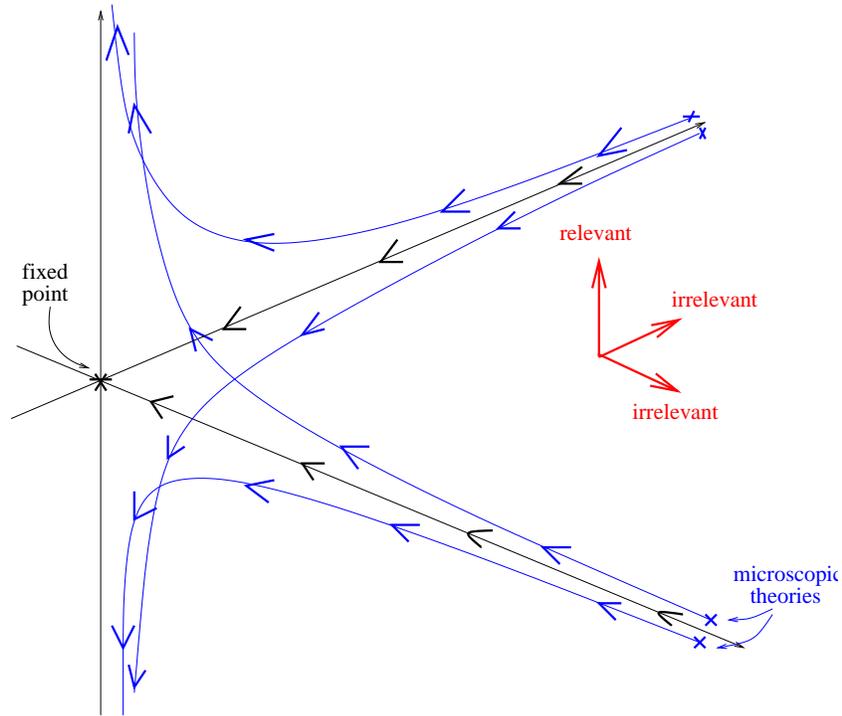}}
\caption{\it Relevant and irrelevant directions in the parameter space. The renormalization
flows are indicated by the blue arrows. Two microscopic theories with
different relevant parameters lead to different large-scale physics, whereas the latter
are the same if the microscopic description differ by irrelevant parameters.}
\label{relirrel}
\end{figure}

"Relevant" and "irrelevant" have actually a precise meaning. One first has to define a fixed point of the 
RGT: this is a theory invariant under the blocking procedure. In principle, a fixed point can describe 
an interacting theory, but usually it describes a free theory.
Once this fixed point $\{\mu^\star\}$ is defined, one can linearize the renormalization group transformations 
around $\{\mu^\star\}$ and define in the space of parameter (see fig.\ref{relirrel}): 
\begin{itemize}
\item Relevant directions: along which the theory goes away from the fixed point. The corresponding
dimensionless parameters become very large when $s\to\infty$ and thus dominate the theory;

\item Irrelevant directions: along which the theory approaches the fixed point. The corresponding
dimensionless parameters become very small when $s\to\infty$ and thus become negligible:
\end{itemize}

One then defines the notion of "universality class" as a group of microscopic theories
having the same macroscopic behaviour. Theories belonging to the same universality class differ by 
irrelevant parameters, and flow towards the same large-distance physics as $s\to\infty$.

How can one finally get physical quantities out of this renormalization procedure?
Such quantities, as magnetic susceptibility or heat capacity, are obtained via correlation functions (see def.4).
Let $G(a)$ be a correlation function defined on the original lattice, homogeneous to a length
to the power $[G]$. After RGTs it can be shown that this correlation function reads, when $s\to\infty$ 
\be
G(sa)\simeq s^{[G]+\eta}G(a),
\ee
where $\eta$ is called the anomalous dimension, arising from thermal fluctuations.
By choosing $s=T_C/(T-T_C)$, which 
goes to infinity when the temperature reaches the critical temperature, one can see that 
this very anomalous dimension can actually be identified with a critical exponent (see def.5) of the theory,
whose prediction is therefore possible with renormalization group methods (the power $[G]$ is a 
natural consequence of the rescaling).

\vspace{.5cm}

This section showed how scale dependence comes into account in the description of a system on 
a lattice. The ideas introduced here go beyond this critical phenomenon and the next section 
shows how they arise in Particle Physics.

\section{Renormalization in Particle Physics - First part}

When one considers elementary processes in Particle Physics, the usual way to proceed is to 
start form a classical description, based on equations of motion, and then look for the quantum
corrections. These quantum corrections involve the exchange of elementary particles, 
or the creation/annihilation of pairs of particles. We consider here an example
taken from Quantum Electrodynamics (QED) which captures the essential features. Among the huge
bibliography explaining theses effects, the reader can look at a good description
for example in \cite{weisskopf}. 

\begin{figure}[ht]
\epsfxsize=10cm
\centerline{\epsffile{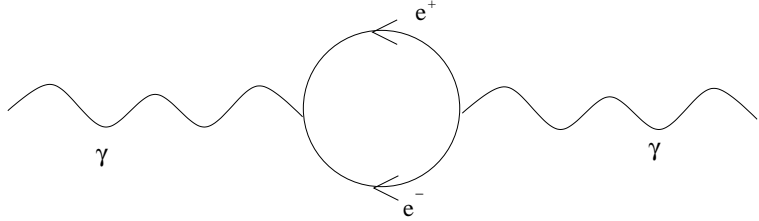}}
\caption{\it Creation/annihilation of a virtual pair electron/positron. Being antiparticles, the
electron and positron have same mass, but opposite charge (and thus electric charge is conserved
at a vertex photon-electron-positron).}
\label{virtualpair}
\end{figure}

In fig.\ref{virtualpair},
a photon propagating creates a pair electron/positron, which finally annihilates to generate another photon.
This process is possible as a consequence of two fundamental properties of relativistic quantum theory:
\begin{itemize}
\item Relativistic aspect: the equivalence mass/energy enables the massless 
photon to "split" into massive electron and positron. For this event to start playing a role, the energy of the
photon should be at least twice the mass energy of the electron, i.e. roughly $10^6$ eV;
\item Quantum aspect: the uncertainty energy/time allows the pair electron/positron to exist, during a time 
proportional to the inverse of their mass. This time is of the order of $10^{-21}$ s.
\end{itemize}

This electron/positron pair is said virtual since it is probabilistic and exists for a very short time only. 
Nevertheless, this quantum effect influences the strength of the 
interaction between electric charges, and therefore contributes to the value of the electron's charge
that is measured. 
 
The computation of the quantum correction to the photon propagation shown in fig.\ref{virtualpair} involves
an integration over all the Fourier modes of the electron/positron. This
integration happens to be divergent if it is done in a straightforward manner. The origin of this divergence is 
the point-like structure of the electron/positron. To avoid this divergence, one has to regularize the
integration, and several techniques exist. For example:

\begin{itemize} 
\item A naive regularization would be to put a cut off in Fourier
space, which is equivalent to give a radius to the electron/positron. But this procedure does not 
respect gauge invariance, which is unacceptable since this would lead to the violation of electric charge
conservation. Therefore this regularization method cannot be used in QED;
\item The Pauli-Villars method, which consists in adding massive particles to the system,
with couplings that are not constrained by physical requirements.
A good choice of these masses and couplings can then cancel the divergences and the additional,
non physical, particles decouple from the dynamics of the system in the limit where their
masses go to infinity.
This method has the drawback to be cumbersome from the computational point of view.
\item The dimensional regularization consists 
in computing the would-be divergent integral in $4-\varepsilon$ dimensions, where $\varepsilon<<1$ 
and "4" stands for 3 space and 1 time dimensions (such integrals can be defined mathematically). 
The would-be divergence then appears as a pole in $\varepsilon$ and
we are left with a regularization dependent electron's charge. 
The reader interested in detailed technical aspects can go to \cite{collins}. Note that this space time
dimension $4-\varepsilon$ has nothing to do with additional dimensions necessary for 
the study of String Theories (the latter extra dimensions are suppose to be real), 
but is just a mathematical trick to give a meaning to the quantum theory. This regularization
is technically more advanced, but is also the simplest one from the computational point of view.
\end{itemize}

Whatever regularization method is used, the essential point is that it {\it necessarily} involves the 
introduction of an energy or mass scale. We recall here that 
length and energy scales are related by Heisenberg inequalities which imply that small
distances correspond to high energies (called ultraviolet - UV - in analogy with the optical spectrum), 
and large distances correspond to low energies (called infrared - IR). In the standards of 
particle accelerators, a typical "low energy"
is of the order of the electron's mass energy ($\simeq 5\times 10^5$ eV)
and a typical "high energy" is of the order of the $Z$ boson's mass energy ($\simeq 9\times 10^{10}$ eV).

Let us continue the discussion with the example of the dimensional regularization.
The next step is to give a meaning to the electron's charge $q$ depending on $\varepsilon$.
This is actually possible when looking at the {\it flow} of $q$ 
with the energy scale $E$ introduced by the regularization: the derivative $dq/dE$
happens to be finite in the limit where $\varepsilon\to 0$.

What we finally obtain is the flow of the electron's charge with an energy scale that was put by hand.
We need then an initial condition chosen at some value of the energy scale, what is done
in the following way: the value of the electron's charge in the limit of low energies is the one measured 
in the laboratory, i.e. in the deep IR region where $E=mc^2$ 
($m$=electron's mass, $c$=speed of light) which is the minimum value for the electron's energy.

\begin{figure}[ht]
\epsfxsize=9cm
\centerline{\epsffile{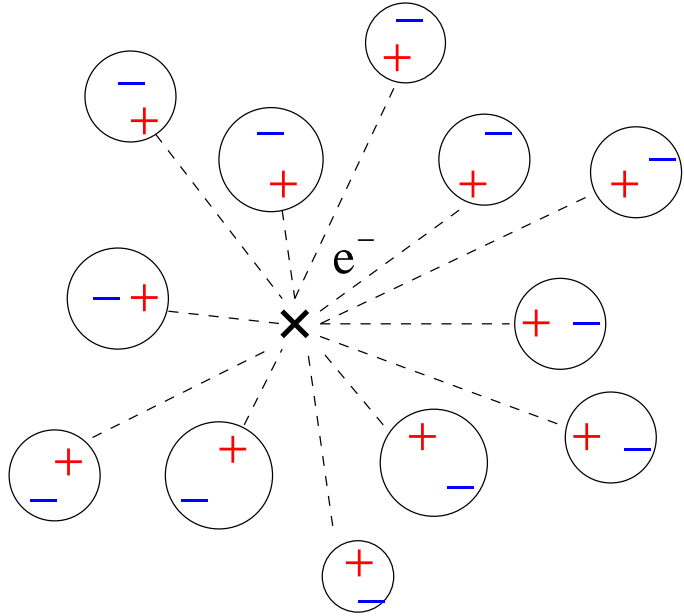}}
\caption{\it An electron polarizes the quantum vacuum made out of virtual electric dipoles electron/positron.
As a consequence, the effective charge that is seen depends on the distance of observation.}
\label{polarization}
\end{figure}

This energy dependence of the electron's charge is not a formal result, but a real 
phenomenon that is observed in particle accelerators: as the energy increases in a collision
of two electrically charged particles, the strength of the interaction (which defines the electron's charge)
increases. The physical interpretation is the following.
The quantum vacuum being full of these virtual electron/positron pairs, an electron put in this
vacuum can polarize it, as shown on fig.\ref{polarization}. As a consequence, 
as one goes away from the electron, one observes a decreasing charge since the electron is screened by the virtual
electric dipoles. The "bare" charge is the one that would be measured at a distance of the
order of the classical electron radius ($\simeq 10^{-15}$m). The physical charge, measured in the laboratory,
is "dressed" by quantum corrections.

\vspace{.2cm}

{\it The Landau pole:} 
In the specific example of the electron's charge running with an energy scale, one actually 
meets a new problem: this charge, increasing with the energy of the physical process, 
seems to diverge at a specific energy, called the "Landau pole". It is true that 
QED is not valid anymore at the
scale of the Landau pole where Quantum Gravity has to be taken into account, but this argument
does not explain why there seems to be a problem in the mathematical structure of the theory. 
It has been argued thought 
\cite{gockler,intqed,gies} that this Landau pole is an artifact coming from the oversimplification 
of the renormalization equations, and that it actually disappears when one takes into account
the evolution of the fermion mass with the occurrence of quantum fluctuations.

\vspace{.2cm}

{\it Renormalizability:} As was described, 
the original parameters that define the bare theory become scale dependent after
regularization and have different values in the IR and in the UV. 
A theory is said to be renormalizable if the number of parameters that have to be redefined 
in this way is finite (QED for example).
In a non-renormalizable theory, one would in principle need to redefine an infinite set of parameters to
give a meaning to the theory. But such a theory can be seen as an effective theory, valid up to 
a certain energy scale only, and thus is not useless. Many of these non renormalizable theories are
actually of great interest.

\vspace{.5cm}

To conclude this section, one can stress that 
the essence of the renormalization procedure lies in the fact that
it is possible to turn a would-be divergence into an energy scale dependence.

\section{Renormalization in Particle Physics - Second part}

We are coming back in this section to the ideas developed in section 2. 
As we have already discussed, the number of interacting degrees of freedom located at the sites of 
a lattice with spacing $a$ becomes huge near a phase transition,
since the correlation length $\xi$ increases. In the limit of 
a diverging correlation length, the ratio $a/\xi$ goes to zero and the degrees of freedom
tend to form a continuous set, what leads to a "field theory". In this situation the 
discreet index labeling a lattice site becomes the continuous coordinate
of a point in space time. This is what happens in Quantum Field Theory, where 
the degrees of freedom are the values of a field at each point $x$ of space time. This field 
is a physical entity that can create or annihilate a particle at the point $x$.

But independently of the discrete/continuous feature, one has a similar situation
as the one in critical phenomena, and thus one can develop the same ideas related to renormalization.
Apparently, the procedure provided by the renormalization group transformations is very different
from the procedure consisting in regulating divergent integrals, but these actually lead to the 
same renormalization flows, as is now discussed. For a review on these methods applied to 
Quantum Field theory, see \cite{janos} and references therein.

The starting point is the path integral (see def.7) 
defining the quantum theory (Richard Feynman, Nobel Prize 1965), the analogue of the partition function:
\be
Z=\int{\cal D}[\phi]\exp\left(\frac{i}{\hbar}S[\phi]\right),
\ee
where $S$ is the classical action (see def.6) 
describing the theory with degrees of freedom $\phi(x)$.
The symbol $\int{\cal D}[\phi]$ stands for the integration 
over all the possible configurations of $\phi$, just as the partition function 
involves a summation over all the microscopic states of the system. 
$\hbar$, the elementary quantum action, plays the role of $k_BT$: it is responsible for 
quantum fluctuations represented here by the oscillating integrand (the complex exponential).

To proceed with the RGTs as these were defined in section 2,  
we write at each point of space time $\phi(x)=\Phi(x)+\tilde\phi(x)$ where $\Phi$ contains the 
IR degrees of freedom and $\tilde\phi$ contains the UV degrees of freedom. 
An easy way to implement this is to go to 
Fourier space and define a momentum scale $k$ such that $\Phi(p)\ne 0$ if $|p|\le k$ and
$\tilde\phi(p)\ne 0$ if $|p|>k$, where $p$ is the momentum, coordinate in Fourier space.
With this decomposition, the original field $\phi$ has the following Fourier components:
\bea
\phi(p)&=&\Phi(p)~~~~\mbox{if}~|p|\le k\nn
\phi(p)&=&\tilde\phi(p)~~~~\mbox{if}~|p|>k,
\eea
and the path integral can be written
\be
Z=\int{\cal D}[\Phi]\int{\cal D}[\tilde\phi]\exp\left(\frac{i}{\hbar}S[\Phi+\tilde\phi]\right),
\ee
where $\int{\cal D}[\tilde\phi]$ stands for the summation over the UV degrees of freedom and
$\int{\cal D}[\Phi]$ the summation over the IR degrees of freedom.
This last equation defines the action $S_k$ of the system observed at length scales $\ge 1/k$:
\bea\label{int2}
\exp\left(\frac{i}{\hbar}S_k[\Phi]\right)
=\int{\cal D}[\tilde\phi]\exp\left(\frac{i}{\hbar}S[\Phi+\tilde\phi]\right).
\eea
Note the similarity with the equation (\ref{int1}) which defines the Hamiltonian $H_s$.
$S_k$ is called the "running action" and contains parameters which take into account all the quantum effects that 
occur at length scales smaller than $1/k$. For very large values of $k$ (large compared to another 
momentum scale of the problem) $S_k$ coincides with the classical action, and as $k\to 0$,
$S_k$ tends to the full "effective action", containing all the quantum fluctuations 
in the system. The effective action is the quantity that is looked for,
and RGTs provide us with an algorithmic way to compute it.

The problem is that the integration (\ref{int2}) is not 
always easy and can only be done perturbatively in most of the interesting cases. 
There is though an interesting point: starting from the cut off $k$, 
an {\it infinitesimal} RGT from $S_k$ to $S_{k-\Delta k}$ leads to
an {\it exact} equation in the limit where $\Delta k\to 0$.
This is the important Wegner-Houghton equation \cite{wegnerhoughton},
which opened the way to the so-called Exact Renormalization Group Equations studies, 
a whole area of research in Particle Physics.

The above procedure has a problem though: it is based on the use of a "sharp cut off" to separate
the IR and UV degrees of freedom, which leads to two problems:

\begin{itemize}
\item If the IR field $\Phi$ is not a constant configuration, the Wegner-Houghton equation
leads to singularities in the limit $\Delta k\to 0$, as a consequence
of the non-differentiability of the sharp cut off;
\item A sharp cut off is not consistent with gauge invariance (see def.8): 
if a gauge symmetry is present in the degrees of freedom
$\phi$, the classification IR/UV field is not valid anymore after a gauge transformation.
\end{itemize}

Another approach, which avoids non-differentiability of the sharp cut off, 
is to introduce a "smooth cut off", providing instead a 
progressive elimination of the UV degrees of freedom, as is done with the Polchinski equation \cite{polchinski}. 
The choice of this smooth cut off is not unique, but it has been argued that the effective action
obtained when $k\to 0$ is independent of this choice 
(for a review, see \cite{wetterich} and references therein). The details of this 
technique will not be discussed here, but the essential point is that the renormalization flows that are obtained
are consistent with those discussed in section 3.
One can understand this in an intuitive way: the would-be divergences in Particle Physics
occur at high energies, or momenta, whereas the physical quantities are obtained in the IR, when the 
energy scale decreases and the parameters get dressed by the quantum corrections. This is the same procedure 
leading to the effective description that is obtained by coarse graining a system. 
Therefore it is expected that both procedures have actually the same physical content.

\vspace{.2cm}

{\it An alternative point of view:}
There is another way to describe the generation of quantum effects in connection with
the concept of scale dependence. 
A detailed description of this procedure is given in \cite{janoskornel}.
The idea is to start from a classical theory containing a very large mass scale, such that quantum
fluctuations are frozen and thus can be neglected compared to the classical description of the system.
As this mass decreases, quantum fluctuations progressively appear in the system and the parameters
tend to their physical values. The interesting point is that it is possible to describe the appearance
of quantum fluctuations by an exact equation, as the Wegner-Houghton or the Polchinski
equation. The advantage though is that this scheme is independent of any cut off procedure, since
it does not deal with a classification of the degrees of freedom in terms of their Fourier modes,
but in terms of their
quantum fluctuations' amplitude. It is consistent with gauge invariance and reproduces the well known
renormalization flows that are obtained by regularizing would-be divergences.

\vspace{.2cm}

An important achievement provided by these different methods is the ressumation of quantum corrections
of different orders in $\hbar$. The method described in section 3, dealing with flows obtained 
after regularizing would-be divergences, is valid order by order in $\hbar$. The renormalized quantities
indeed have to be defined at every order in the perturbative expansion 
in $\hbar$. Instead, the exact renormalization methods are not perturbative and take into account 
all the orders in $\hbar$. It should not be forgotten though, that in order to solve these
exact renormalization equations, one needs to make assumptions on the functional dependence of the 
running action (degrees of freedom dependence). These assumptions are usually based on what is called
the gradient expansion, an expansion of the action in powers of the derivatives of the 
field. This approximation is valid in the IR since it assumes low momenta of the physical processes.
There are other assumptions that can be made, and the relevance of each of these approximations
depends on the system that is studied.

\section{Conclusion}

Renormalization provides a common framework to critical phenomena and Particle Physics,
where scale dependence plays an important role in describing Physics in a consistent way.
Renormalization flows of parameters defining a system relate microscopic details to
macroscopic behaviours. In this context the IR theory gives an effective description
taking into account the UV dynamics dressed by quantum corrections. 

Note that the concept of renormalization is in principle independent of would-be divergences.
It is the presence of these divergences which enforces the use of renormalization. The toy model QED
in 2 space dimensions is an example where no divergences occur but the flow of the parameters 
with the amplitude of the quantum corrections can be studied \cite{qed2+1}. In such a theory, 
physical quantities can be expressed in terms of the bare as well as the dressed parameters.

Let us make here a comment on supersymmetric theories, which have been introduced in order to cancel
dominant divergences. In these models, each bosonic degree of freedom has a 
fermionic partner and vice versa. This feature has the effect to cancel some of the quantum corrections and, 
as a result, some of the bare parameters do not get dressed after quantization. Supersymmetric particles,
though, have not been observed experimentally yet, but might be confirmed by the next generation 
of particle accelerators. An old but classic and pedagogical review of supersymmetry is given in
\cite{sohnius}.

Finally, one can consider that every theory is an effective theory, resulting from the elimination
of processes occurring at higher energies, or at smaller lengths. The ultimate theory, from which 
everything is then generated is called the "Theory of Everything" and is certainly not ready to be found...
But the essential point is that, looking for this Theory of Everything, one arrives at many exciting and
challenging achievements.

\section*{Appendix}

\nin 1 eV$\simeq 1.6\times 10^{-19}$ J~~~~
$\hbar\simeq 10^{-34}$ m$^2$ kg s$^{-1}$~~~~
$k_B\simeq 1.4\times 10^{-23}$ m$^2$ kg s$^{-2}$ K$^{-1}$

\vspace{.2cm}

\nin{\bf def.1} \underline{$2^{nd}$ order phase transition}:
For such a transition, first derivatives of the Gibbs free energy are contiunous, 
as the entropy or volume per unit mole, such that there is no latent heat and the two phases do not coexist.
The discontinuous physical quantities are given by second derivatives of the Gibbs free energy, as the heat 
capacity or the thermal expansivity.

\vspace{.2cm}

\nin{\bf def.2} \underline{Hamiltonian}:
Function of the degrees of freedom, whose values are the energies of the system.
In a quantum description, degrees of freedom are replaced by operators, such that the Hamiltonian is an
operator, whose eigen values are the energies of the system.

\vspace{.2cm}

\nin{\bf def.3} \underline{Partition function}: 
Sum over the microscopic states of a system of the 
Boltzmann factors $\exp(-H/k_BT)$, where $H$=Hamiltonian of a given configuration, $k_B$=Boltzmann constant and
$T$=temperature. The computation of the partition function, as a function of $T$, 
leads to the complete knowledge of the
physical properties of the system in contact with a heat source at temperature $T$.

\vspace{.2cm}

\nin{\bf def.4} \underline{Correlation function}: 
Thermal (or quantum) fluctuations at different points of a system are correlated, and correlation functions
measure how these correlations depend on the distance between the points where fluctuations are looked at.
The knowledge of these correlation functions leads to the prediction of physical properties of the system.

\vspace{.2cm}

\nin{\bf def.5} \underline{Critical exponent}: 
Near a $2^{nd}$ order phase transition, several physical quantities diverge as 
the temperature $T$ reaches the critical temperature $T_C$. This divergence can be expressed as a power law
of the form $1/t^\alpha$, where $t=(T/T_C-1)\to 0$ and $\alpha$ is the critical exponent. 

\vspace{.2cm}

\nin{\bf def.6} \underline{Action}: A classical system follows a trajectory which is found by minimizing
a function of its degrees of freedom. This function is the action and is homogeneous to Energy$\times$Time.
In the case of field theory, $S$ is a "functional" = a function of a continuous set of variables.

\vspace{.2cm}

\nin{\bf def.7} \underline{Path integral}: A quantum system does not follow a trajectory: its degrees 
of freedom can take any values, due to quantum fluctuations. The quantum states of the system are randomly 
distributed around the would-be classical trajectory, and the path integral is the sum over every "quantum
trajectory" (not necessarily differentiable),
each of which is attributed the complex weight $\exp(iS/\hbar)$, where $S$=action corresponding
to a given quantum trajectory and $\hbar$=Plank constant$/2\pi$. 
The computation of the path integral, as a function of
the source which generates a (classical) state of the system, leads to the complete knowledge of the
latter.

\vspace{.2cm}
                                                                                                                      
\nin{\bf def.8} \underline{Gauge invariance}: The electromagnetic field can be expressed in
terms of potentials, which are not unique: specific redefinitions of the potentials should let the electromagnetic
field unchanged, which is called gauge invariance. The quantization of the electromagnetic field is
based on these potentials, such that gauge invariance must be preserved throughout the quantization procedure,
for the physical quantities to be gauge invariant.

\end{document}